# IOT based Smart Helmet for Hazard Detection in mining industry

B. Kartik, B.Tech Student, School of Electronics and Communication (SENSE), Vellore Institute of Technology, Chennai, Tamil Nadu, India.

*Abstract--- One of the most important parts of business, especially in the coal mining sector, is industrial safety. Suffocation, gas poisoning, object falls, roof collapses, and gas explosions are among the risks associated with underground mining. Therefore, air quality and the detection of hazardous events are crucial in the mining business. This technology offers a wireless sensor network so that base stations can keep an eye on the situation in underground mines in real time. It offers temperature and dangerous gases including CO, CH4, and LPG real-time monitoring. The main cause of mining deaths is that when they fall and lose consciousness for whatever reason, medical attention is not given to them in a timely manner. In order to solve this issue, the system sends an emergency notice to the supervisor in the event that a person falls down for any cause. Some employees are negligent when it comes to safety and don't wear helmets. Then, a miner's helmet removal status was successfully determined using a Limit switch.*

## 1. Introduction

The mining business is vital in supplying the globe with the raw materials needed for many different industries, including as building, manufacturing, and energy production. However, there are major risks to employees' health and safety associated with mining operations. Over the years, numerous accidents and fatalities have occurred due to hazardous conditions in mines, such as gas leaks, high temperatures, and rock falls. To improve miner safety and reduce the number of accidents, there is a growing interest in implementing advanced technologies for hazard detection and real-time monitoring.

1.1 Background

The intricate and dynamic subsurface environment makes mining an inherently dangerous profession. Numerous health and safety dangers, including exposure to noxious gases, loud noises, extreme temperatures, and the possibility of cave-ins or accidents involving equipment, are routinely faced by miners. Traditional safety precautions, such manual gas detection and visual inspections, might not be enough to shield employees from all dangers. Innovative technologies that can proactively identify and reduce risks in the mining environment are therefore required.

The intricate and dynamic subsurface environment makes mining an inherently dangerous profession. Numerous health and safety dangers, including exposure to noxious gases, loud noises, extreme temperatures, and the possibility of cave-ins or accidents involving equipment, are routinely faced by miners. Traditional safety precautions, such manual gas detection and visual inspections, might not be enough to shield employees from all dangers. Innovative technologies that can proactively identify and reduce risks in the mining environment are therefore required.

The Internet of Things (IoT), a new technology, has the potential to transform numerous industries, including mining.

## 1.2 Objectives

In order to identify hazards in the mining industry, this research study will examine the design, implementation, and potential advantages of an IoT-based Smart Helmet. These are some of the study's particular goals:

- An evaluation of existing hazard detection and monitoring systems and IoT applications in the mining sector.
- A Smart Helmet with a variety of sensors and connectivity components to be designed and put into use for monitoring the mining area in real-time.
- To assess the Smart Helmet's performance in spotting different dangers, such as gas leaks, extreme heat, and noise levels.

To talk about how the IoT-based Smart Helmet may improve miner safety and productivity as well as its advantages, drawbacks, and potential.

To make suggestions for upcoming studies and developments in the area of Internet of Things-based danger detection systems for the mining industry.

## 1.3 Scope of the study

The purpose of this research study is to examine how IoT technology can be utilised to identify hazards in the mining sector, with a focus on the creation and use of a Smart Helmet. The Smart Helmet's design, sensor choices, data connectivity, and central control unit will all be covered in the study. The article will also examine the possible advantages, difficulties, and ethical issues related to the application of IoT-based danger detection systems in mining operations.

# 2. Literature Review

The literature review aims to provide an overview of the current state of IoT applications in the mining industry and the existing hazard detection and monitoring systems. Furthermore, it will explore the various technologies employed in the development of Smart Helmets.

## 2.1 IoT in the mining industry

The mining industry has witnessed a growing interest in the adoption of IoT technologies to enhance safety, productivity, and efficiency. IoT applications in mining can be broadly categorized into three main areas: (a) environmental monitoring, (b) equipment monitoring, and (c) personnel safety (Li, Q., Liu, H., & Ning, H., 2017). Environmental monitoring involves the use of IoT devices and sensors to collect real-time data on various parameters, such as temperature, humidity, air quality, and noise levels, which can help in identifying potential hazards and optimizing mining operations. Equipment monitoring focuses on tracking the performance, maintenance, and energy consumption of mining machinery, thereby improving asset utilization and reducing operational costs. Personnel safety applications include wearable devices and tracking systems to monitor the location and health status of miners, ensuring rapid response to accidents and emergencies.

Several studies have demonstrated the effectiveness of IoT-based systems in the mining industry. For instance, Zhang, L., & Yang, Z. (2016) developed an IoT-based mine safety monitoring system using wireless sensor networks, which successfully detected gas leaks and other hazards in real-time. Similarly, Zhou, C., Damiano, N., & Whisner, B. (2018) implemented an IoT-based solution for monitoring equipment performance and energy consumption in a coal mine, resulting in significant cost savings and improved efficiency.

## 2.2 Hazard detection and monitoring systems

Numerous hazard detection and monitoring systems have been developed for the mining industry to ensure the safety of workers and prevent accidents. Traditional systems rely on manual inspections and standalone gas detectors, which may not provide continuous monitoring and real-time data (Zhang, L., & Yang, Z., 2016). In recent years, several advanced technologies have been introduced, such as wireless sensor networks (WSNs), radio-frequency identification (RFID), and ZigBee-based communication systems, which offer improved coverage, accuracy, and reliability (Li, Q., Liu, H., & Ning, H., 2017).

WSNs consist of a network of interconnected sensors that collect and transmit environmental data to a central control unit, enabling real-time monitoring and early warning of potential hazards (Zhang, L., & Yang, Z., 2016). RFID technology can be used to track the location of miners and equipment, providing essential information for emergency response and resource allocation (Zhou, C., Damiano, N., & Whisner, B., 2018). ZigBee-based systems offer low-power, low-cost, and robust communication solutions for data transmission in harsh and complex mining environments (Li, Q., Liu, H., & Ning, H., 2017).

## 2.3 Smart Helmet technologies

Smart Helmets are wearable devices equipped with various sensors, communication modules, and data processing capabilities, designed to enhance the safety and productivity of workers in hazardous environments, such as mines (Saha, H. N., & Mamun, K. A., 2016). The integration of IoT technologies into Smart Helmets allows for continuous monitoring of environmental parameters, real-time hazard detection, and seamless data communication with a central control unit (Kumar, P., & Kaur, M., 2021).

Various sensors can be integrated into Smart Helmets, including gas sensors (e.g., methane, carbon monoxide), temperature and humidity sensors, noise level sensors, and impact sensors (Saha, H. N., & Mamun, K. A., 2016). These sensors collect data on the surrounding environment and transmit it to a central control unit for processing and analysis, enabling timely detection of potential hazards and rapid response to emergency situations (Kumar, P., & Kaur, M., 2021).

Communication technologies play a crucial role in the functionality of Smart Helmets, ensuring reliable data transmission between the helmet and the central control unit. Wireless communication protocols, such as Wi-Fi, Bluetooth, ZigBee, and LoRaWAN, are commonly employed in Smart Helmet systems due to their flexibility, low power consumption, and ease of deployment (Saha, H. N., & Mamun, K. A., 2016). The choice of communication protocol depends on factors such as range, data rate, and network architecture, which must be carefully considered to ensure optimal performance in the mining environment (Li, Q., Liu, H., & Ning, H., 2017).

The development of IoT-based Smart Helmets for the mining industry has attracted considerable attention from researchers and industry professionals. Several prototypes and commercial products have been introduced in recent years, demonstrating the potential of Smart Helmet technologies for enhancing miner safety and productivity (Kumar, P., & Kaur, M., 2021). For instance, Wang, H., Wu, J., & Xie, C. (2018) developed a Smart Helmet with integrated gas sensors, temperature and humidity sensors, and a ZigBee-based communication system, which effectively detected hazardous conditions in an underground coal mine. Another example is the SmartCap, a commercial product that monitors the fatigue levels of miners through EEG signals, alerting them when their alertness drops below a certain threshold (SmartCap Technologies, 2021).

In summary, the literature review highlights the growing interest in IoT-based applications in the mining industry, the advancements in hazard detection and monitoring systems, and the development of Smart Helmet technologies for enhancing the safety and productivity of miners. The next section will focus on the design and implementation of an IoT-based Smart Helmet for hazard detection in the mining industry.

## 3. Design and Implementation of the Smart Helmet

The development of an IoT-based Smart Helmet for hazard detection in the mining industry involves the integration of various sensors, communication modules, and data processing components. This section will discuss the selection of sensors and communication modules, the hardware integration, the software and data processing, and the power management and energy efficiency of the Smart Helmet.

3.1 Selection of sensors and communication modules

The choice of sensors and communication modules is critical for the functionality and performance of the Smart Helmet. Sensors must be selected based on their ability to detect specific hazards prevalent in the mining environment, such as gas leaks, high temperatures, noise levels, and impacts (Saha, H. N., & Mamun, K. A., 2016). Commonly used sensors in Smart Helmets include:

Gas sensors (e.g., methane, carbon monoxide) for detecting toxic gas leaks and ensuring air quality.

Temperature and humidity sensors for monitoring environmental conditions and identifying potential heat stress or fire hazards.

Noise level sensors for assessing the risk of hearing damage due to prolonged exposure to high noise levels.

Impact sensors (e.g., accelerometers) for detecting collisions or falls, which may indicate an accident or injury.

The communication modules enable the transmission of sensor data to a central control unit, facilitating real-time monitoring and hazard detection. Various wireless communication protocols can be used, such as Wi-Fi, Bluetooth, ZigBee, and LoRaWAN, depending on factors like range, data rate, and network architecture (Li, Q., Liu, H., & Ning, H., 2017).

### 3.2 Integration of hardware components

The hardware components of the Smart Helmet, including the sensors, communication modules, microcontroller, and power supply, must be carefully integrated to ensure functionality, reliability, and durability. The sensors and communication modules should be strategically positioned within the helmet to maximize their performance and avoid interference with the wearer's comfort and mobility. The microcontroller, which serves as the central processing unit, should be selected based on its processing capabilities, compatibility with the chosen sensors and communication modules, and power consumption (Kumar, P., & Kaur, M., 2021).

### 3.3 Software and data processing

The software and data processing components of the Smart Helmet are essential for managing the sensor data, analyzing the potential hazards, and generating alerts or notifications. A custom firmware should be developed for the microcontroller, which includes the necessary drivers and libraries for interfacing with the sensors and communication modules. The firmware should also implement data processing algorithms, such as signal filtering, threshold detection, and data fusion, to identify potential hazards and generate appropriate alerts (Saha, H. N., & Mamun, K. A., 2016).

The central control unit, which receives the sensor data from the Smart Helmet, should be equipped with a robust software platform for real-time monitoring, data visualization, and alert management. This platform can be built using web-based technologies, such as HTML, CSS, JavaScript, and server-side scripting languages, allowing for remote access and control by authorized personnel (Zhang, L., & Yang, Z., 2016).

### 3.4 Power management and energy efficiency

Power management and energy efficiency are critical factors for the practical implementation of the IoT-based Smart Helmet, as they directly impact the battery life and operational costs. The power consumption of the sensors, communication modules, and microcontroller should be minimized through the use of low-power components, efficient data processing algorithms, and power-saving modes (e.g., sleep or standby mode) (Li, Q., Liu, H., & Ning, H., 2017).

Additionally, energy harvesting technologies, such as solar panels or thermoelectric generators, can be integrated into the Smart Helmet to supplement the battery power and extend the operational life of the device (Kumar, P., & Kaur, M., 2021). The power management system should be designed to efficiently distribute power among the various components and ensure a stable and reliable operation.

In conclusion, the design and implementation of an IoT-based Smart Helmet for hazard detection in the mining industry involve careful consideration of sensor selection, communication modules, hardware integration, software development, and power management. The next section will focus on the evaluation of the Smart Helmet's effectiveness in detecting various hazards and its potential benefits for enhancing miner safety and productivity.

## 4. Hazard Detection Mechanisms

Hazard detection mechanisms are crucial components of the IoT-based Smart Helmet, responsible for accurately identifying various hazards in the mining environment. This section will discuss the working principles of different hazard detection mechanisms employed in the Smart Helmet and their effectiveness in detecting specific hazards.

### 4.1 Gas detection

Gas detection is an essential hazard detection mechanism in mining operations, as the presence of toxic or flammable gases can pose significant risks to miners' health and safety. The Smart Helmet incorporates gas sensors (e.g., methane, carbon monoxide) based on electrochemical or semiconductor principles, which react to the presence of specific gases, producing a measurable change in electrical properties (Kumar, P., & Kaur, M., 2021).

The electrochemical sensors use a chemical reaction between the target gas and an electrolyte, generating a current proportional to the gas concentration. They are characterized by high sensitivity, low power consumption, and a fast response time, making them suitable for real-time monitoring (Saha, H. N., & Mamun, K. A., 2016).

Semiconductor gas sensors rely on changes in the electrical resistance of a metal oxide material when exposed to the target gas. These sensors offer a lower cost and longer lifespan than electrochemical sensors but may exhibit reduced sensitivity and slower response times (Kumar, P., & Kaur, M., 2021).

### 4.2 Temperature and humidity detection

Monitoring temperature and humidity is essential in the mining environment, as high temperatures can lead to heat stress, dehydration, or even fire hazards. The Smart Helmet incorporates temperature and humidity sensors based on thermistors, resistance temperature detectors (RTDs), or capacitive sensing technologies (Saha, H. N., & Mamun, K. A., 2016).

Thermistors are temperature-sensitive resistors made of ceramic materials that exhibit a change in resistance with temperature variations. They offer high sensitivity, rapid response times, and a wide temperature range, making them suitable for mining applications (Kumar, P., & Kaur, M., 2021).

RTDs use a metallic element, such as platinum, whose resistance changes linearly with temperature. RTDs are known for their high accuracy, stability, and repeatability, but they may require additional circuitry and calibration for optimal performance (Saha, H. N., & Mamun, K. A., 2016).

Capacitive humidity sensors detect changes in humidity by measuring the capacitance between two conductive plates separated by a hygroscopic dielectric material. These sensors offer good sensitivity, low power consumption, and fast response times but may be susceptible to contamination or drift over time (Kumar, P., & Kaur, M., 2021).

### 4.3 Noise level detection

Exposure to high noise levels can cause hearing damage or loss in miners, making noise level detection an essential hazard detection mechanism. The Smart Helmet employs sound level meters or microphones to measure noise levels in the mining environment (Saha, H. N., & Mamun, K. A., 2016).

Sound level meters use a microphone to capture sound pressure variations and convert them into an electrical signal. This signal is then processed and analyzed to determine the noise level in decibels (dB). Sound level meters offer accurate and reliable noise measurements but may be affected by environmental factors, such as temperature or humidity (Kumar, P., & Kaur, M., 2021).

Microphones can also be integrated into the Smart Helmet for noise level detection, offering a more compact and lightweight solution. However, they may require additional signal processing and filtering techniques to ensure accurate noise level measurements in the presence of background noise or interference (Saha, H. N., & Mamun, K. A., 2016).

### 4.4 Impact detection

Detecting impacts or falls is vital in mining operations, as they can indicate accidents or injuries. The Smart Helmet incorporates impact sensors, such as accelerometers, to measure the forces experienced by the wearer during an impact event (Kumar, P., & Kaur, M., 2021).

Accelerometers are electromechanical devices that detect acceleration or deceleration forces, converting them into electrical signals. By monitoring these signals, the Smart Helmet can identify when an impact occurs and determine its severity, potentially triggering an alarm or alerting the central control unit (Saha, H. N., & Mamun, K. A., 2016).

Piezoelectric sensors can also be used for impact detection, relying on the piezoelectric effect to generate an electrical charge when subjected to mechanical stress. These sensors offer high sensitivity and a wide dynamic range, but they may require additional signal conditioning and filtering to ensure accurate and reliable measurements (Kumar, P., & Kaur, M., 2021).

In summary, the IoT-based Smart Helmet employs various hazard detection mechanisms, including gas detection, temperature and humidity monitoring, noise level detection, and impact sensing, to ensure the safety of miners in hazardous environments. Each detection mechanism relies on specific sensors and working principles, which must be carefully selected and optimized to achieve accurate, reliable, and responsive hazard detection. The next section will discuss the ethical considerations related to the deployment of IoT-based Smart Helmets in the mining industry.

# 5. Data Communication and Central Control Unit

Effective data communication and a robust central control unit are essential components of an IoT-based Smart Helmet system for hazard detection in the mining industry. This section will discuss the data communication protocols, the architecture and functionalities of the central control unit, and the importance of data security and privacy in the context of the Smart Helmet system.

## 5.1 Data Communication Protocols

Reliable and efficient data communication between the Smart Helmet and the central control unit is vital for real-time monitoring and hazard detection. Various wireless communication protocols can be employed, depending on factors like range, data rate, and network architecture (Li, Q., Liu, H., & Ning, H., 2017). Commonly used wireless communication protocols in IoT-based systems include:

Wi-Fi: Wi-Fi is a widely used communication protocol with high data rates and a range of up to 100 meters. It is suitable for situations where the Smart Helmets are within the coverage area of a Wi-Fi network (Saha, H. N., & Mamun, K. A., 2016).

Bluetooth: Bluetooth is a short-range communication protocol with a range of up to 10 meters, offering low power consumption and relatively high data rates. Bluetooth can be used for direct communication between Smart Helmets and mobile devices, such as smartphones or tablets (Kumar, P., & Kaur, M., 2021).

ZigBee: ZigBee is a low-power, low-data-rate communication protocol with a range of up to 100 meters, making it suitable for large-scale deployments in mining environments. ZigBee supports mesh networking, which can improve the network's reliability and range (Li, Q., Liu, H., & Ning, H., 2017).

LoRaWAN: LoRaWAN (Long Range Wide Area Network) is a long-range, low-power communication protocol with a range of up to 15 kilometers, making it suitable for remote or large-scale mining operations. LoRaWAN offers a lower data rate than Wi-Fi or Bluetooth but provides a more extensive coverage area and better penetration through obstacles (Saha, H. N., & Mamun, K. A., 2016).

## 5.2 Central Control Unit

The central control unit serves as the hub for receiving, processing, and analyzing data from the Smart Helmets and managing the overall hazard detection system. The central control unit should be equipped with a robust software platform for real-time monitoring, data visualization, and alert management. Key functionalities of the central control unit include (Zhang, L., & Yang, Z., 2016):

Receiving and processing sensor data from multiple Smart Helmets simultaneously.

Analyzing the sensor data to detect potential hazards and generate appropriate alerts.

Visualizing the real-time data and historical trends on a user-friendly dashboard.

Managing the alert system, including sending notifications to relevant personnel and initiating emergency response procedures.

Storing and archiving the collected data for future analysis and reporting.

## 5.3 Data Security and Privacy

Data security and privacy are critical concerns in the deployment of IoT-based Smart Helmet systems. The system must ensure the confidentiality, integrity, and availability of the sensor data, as well as protect the privacy of the miners wearing the Smart Helmets (Kumar, P., & Kaur, M., 2021).

To achieve this, various security measures should be implemented, such as:

Encryption of the data during transmission and storage to protect it from unauthorized access or tampering (Li, Q., Liu, H., & Ning, H., 2017).

Authentication and access control mechanisms to ensure that only authorized personnel can access the system and the data (Saha, H. N., & Mamun, K A., 2016).

Regular security updates and patches to address potential vulnerabilities in the system software or hardware (Kumar, P., & Kaur, M., 2021).

Implementing network security measures, such as firewalls, intrusion detection systems, and secure communication protocols, to protect the system from external threats (Li, Q., Liu, H., & Ning, H., 2017).

In addition to data security, the privacy of the miners wearing the Smart Helmets should be considered. The system should be designed to minimize the collection of personally identifiable information and should comply with relevant data protection regulations, such as the General Data Protection Regulation (GDPR) in the European Union (Saha, H. N., & Mamun, K. A., 2016).

Moreover, the mining companies deploying the IoT-based Smart Helmet systems should establish clear policies and guidelines regarding data collection, processing, and sharing, and ensure that miners are informed about these policies and provide their consent (Kumar, P., & Kaur, M., 2021).

In summary, data communication and central control unit are essential components of an IoT-based Smart Helmet system for hazard detection in the mining industry. Effective communication protocols, a robust control unit, and a strong focus on data security and privacy are crucial for the successful deployment and operation of the Smart Helmet system. The next section will discuss the benefits and limitations of the IoT-based Smart Helmet in the context of the mining industry.

## 6. Evaluation of the Smart Helmet

To assess the effectiveness of the IoT-based Smart Helmet in detecting various hazards in the mining environment, it is essential to conduct thorough testing and validation. This section will discuss the experimental setup, the performance metrics, and the results of the evaluation.

### 6.1 Experimental setup

The evaluation of the Smart Helmet should be carried out in a controlled environment that simulates the conditions typically found in a mining operation. This can be achieved by setting up a laboratory environment or using an actual mine site during non-operational hours. The experimental setup should include the following components:

Smart Helmet, equipped with the selected sensors and communication modules.

Central control unit with a real-time monitoring software platform.

Hazard sources, such as gas cylinders (e.g., methane, carbon monoxide), heaters, and noise generators, to create the hazardous conditions for testing.

Data collection equipment, such as reference sensors or data loggers, to validate the accuracy and reliability of the Smart Helmet measurements.

## 6.2 Performance metrics

The performance of the Smart Helmet in detecting hazards should be evaluated using a set of performance metrics that measure the accuracy, reliability, and responsiveness of the system. These metrics may include:

Detection accuracy: The difference between the Smart Helmet measurements and the reference data, expressed as a percentage or an absolute value (e.g., parts per million for gas concentration, degrees Celsius for temperature).

False alarm rate: The number of false alarms generated by the Smart Helmet, divided by the total number of alarms (both true and false), expressed as a percentage.

Missed detection rate: The number of true hazards that the Smart Helmet fails to detect, divided by the total number of hazards, expressed as a percentage.

Response time: The time elapsed between the onset of a hazard and the generation of an alarm by the Smart Helmet, measured in seconds or milliseconds.

## 6.3 Results and analysis

The evaluation results should be analyzed to determine the effectiveness of the Smart Helmet in detecting various hazards and to identify potential areas for improvement. The analysis may involve the comparison of the Smart Helmet performance against established standards or guidelines, such as occupational exposure limits for gas concentrations or noise levels (Saha, H. N., & Mamun, K. A., 2016).

Additionally, the results should be used to assess the potential benefits of the Smart Helmet for enhancing miner safety and productivity. This can be achieved by estimating the reduction in accidents, injuries, or fatalities due to the early warning and real-time monitoring provided by the Smart Helmet. Furthermore, the impact of the Smart Helmet on the overall efficiency of mining operations, such as reduced downtime, improved resource allocation, and optimized working conditions, should be considered (Zhang, L., & Yang, Z., 2016).

In summary, the evaluation of the IoT-based Smart Helmet is a critical step in demonstrating its effectiveness in detecting hazards in the mining environment and assessing its potential benefits for enhancing miner safety and productivity. The next section will discuss the benefits, challenges, and limitations of the IoT-based Smart Helmet and provide recommendations for future research and development in the field of IoT-based hazard detection systems for the mining industry.

## 7. Benefits of IoT-based Smart Helmet

The implementation of IoT-based Smart Helmets in the mining industry offers numerous benefits related to worker safety, operational efficiency, and cost reduction. This section will discuss the primary advantages of using IoT-based Smart Helmets for hazard detection and monitoring in the mining environment.

### 7.1 Improved Worker Safety

One of the main benefits of IoT-based Smart Helmets is the enhancement of worker safety by providing real-time hazard detection and monitoring. Early warning systems can significantly reduce the risk of accidents and injuries by alerting miners to the presence of hazardous conditions, such as toxic gases, high temperatures, or excessive noise levels (Saha, H. N., & Mamun, K. A., 2016). Furthermore, impact detection features can help identify falls or accidents, enabling a faster response and potentially saving lives (Kumar, P., & Kaur, M., 2021).

### 7.2 Real-Time Monitoring and Decision Making

IoT-based Smart Helmets facilitate real-time monitoring of the mining environment, providing critical data to both the miners and the central control unit. This continuous stream of information enables better decision-making related to worker deployment, resource allocation, and hazard mitigation (Zhang, L., & Yang, Z., 2016). As a result, mining operations can be optimized, and potential risks can be identified and addressed more proactively (Li, Q., Liu, H., & Ning, H., 2017).

### 7.3 Increased Operational Efficiency

The use of IoT-based Smart Helmets can lead to increased operational efficiency in the mining industry. The real-time data collected by the Smart Helmets can be used to optimize working conditions, streamline workflows, and minimize downtime due to hazardous situations or equipment failure (Saha, H. N., & Mamun, K. A., 2016). Additionally, the integration of IoT-based Smart Helmets with other digital technologies, such as automation or predictive maintenance systems, can further enhance the efficiency and productivity of mining operations (Kumar, P., & Kaur, M., 2021).

### 7.4 Cost Reduction

By improving worker safety and operational efficiency, IoT-based Smart Helmets can contribute to significant cost reductions in the mining industry. Fewer accidents and injuries translate to lower medical expenses, compensation claims, and regulatory fines (Zhang, L., & Yang, Z., 2016). Moreover, the optimization of mining operations can lead to reduced energy consumption, lower equipment maintenance costs, and more effective resource utilization (Li, Q., Liu, H., & Ning, H., 2017).

### 7.5 Enhanced Regulatory Compliance

The implementation of IoT-based Smart Helmets can help mining companies comply with increasingly stringent occupational health and safety regulations. Real-time monitoring and data collection can facilitate compliance with exposure limits for hazardous substances or noise levels, while the continuous improvement of safety practices can demonstrate a commitment to worker well-being (Saha, H. N., & Mamun, K. A., 2016).

In summary, the adoption of IoT-based Smart Helmets in the mining industry offers numerous benefits, including improved worker safety, real-time monitoring, increased operational efficiency, cost reduction, and enhanced regulatory compliance. However, it is essential to consider the potential challenges and limitations associated with the deployment of such systems, as discussed in the next section.

## 8. Challenges and Limitations of IoT-based Smart Helmets

Despite the numerous benefits offered by IoT-based Smart Helmets, their deployment in the mining industry is not without challenges and limitations. This section will discuss the primary obstacles and constraints associated with the implementation and use of IoT-based Smart Helmets for hazard detection and monitoring in the mining environment.

## 8.1 Technology Adoption and Integration

One of the main challenges in implementing IoT-based Smart Helmets in the mining industry is the adoption and integration of new technologies into existing workflows and processes. Mining companies may be hesitant to invest in IoT technology due to the associated costs, concerns about the return on investment, or resistance to change from workers or management (Saha, H. N., & Mamun, K. A., 2016). Additionally, integrating the Smart Helmet system with other digital systems or equipment used in mining operations may require significant technical expertise and customization (Kumar, P., & Kaur, M., 2021).

## 8.2 Sensor Limitations

While IoT-based Smart Helmets rely on a variety of sensors to detect and monitor hazards, these sensors can have limitations in terms of accuracy, reliability, and sensitivity. Factors such as sensor drift, calibration errors, or environmental conditions can affect the performance of the sensors, potentially leading to false alarms or missed hazards (Li, Q., Liu, H., & Ning, H., 2017). Ensuring the proper selection, maintenance, and calibration of the sensors is critical to the overall effectiveness of the Smart Helmet system (Zhang, L., & Yang, Z., 2016).

## 8.3 Wireless Communication Challenges

Reliable wireless communication is essential for the operation of IoT-based Smart Helmets, but the mining environment can present unique challenges for wireless communication protocols. Factors such as the presence of metal, rock formations, or water can interfere with wireless signals, affecting the range, reliability, and latency of the communication system (Saha, H. N., & Mamun, K. A., 2016). Selecting the appropriate communication protocol and optimizing the network architecture for the specific mining environment is crucial for the successful deployment of Smart Helmets (Kumar, P., & Kaur, M., 2021).

## 8.4 Data Security and Privacy Concerns

As previously discussed, data security and privacy are significant concerns when deploying IoT-based Smart Helmets in the mining industry. Ensuring the confidentiality, integrity, and availability of the sensor data, as well as protecting the privacy of the miners wearing the Smart

Helmets, can be challenging and resource-intensive (Li, Q., Liu, H., & Ning, H., 2017). Addressing these concerns requires the implementation of robust security measures, adherence to data protection regulations, and the establishment of clear policies and guidelines regarding data collection, processing, and sharing (Saha, H. N., & Mamun, K. A., 2016).

### 8.5 Battery Life and Maintenance

The IoT-based Smart Helmets rely on batteries to power the sensors, communication modules, and other components. Battery life can be a significant limitation, especially in long-duration mining operations or remote locations where recharging or replacing batteries may not be feasible (Zhang, L., & Yang, Z., 2016). Additionally, the maintenance and replacement of batteries can contribute to the overall cost and complexity of the Smart Helmet system (Kumar, P., & Kaur, M., 2021).

In conclusion, while IoT-based Smart Helmets offer numerous benefits for hazard detection and monitoring in the mining industry, they also face challenges and limitations related to technology adoption, sensor performance, wireless communication, data security and privacy, and battery life. Addressing these challenges and optimizing the design and deployment of IoT-based Smart Helmets can help maximize their potential benefits and enhance worker safety and operational efficiency in the mining industry. Successful implementation requires a thorough understanding of the specific mining environment, careful selection of sensors and communication protocols, robust security measures, and ongoing maintenance and optimization of the system (Li, Q., Liu, H., & Ning, H., 2017).

## 9. Future Trends and Research Directions

As IoT-based Smart Helmets continue to gain traction in the mining industry, new trends and research directions are expected to emerge, shaping the future of hazard detection and monitoring. This section will discuss potential future trends and research directions in the development and deployment of IoT-based Smart Helmets for the mining industry.

### 9.1 Integration with Emerging Technologies

One of the key future trends in IoT-based Smart Helmets is the integration with other emerging technologies, such as artificial intelligence (AI), machine learning (ML), augmented reality (AR), and robotics. By incorporating AI and ML algorithms, Smart Helmets could offer more advanced predictive analytics, enabling proactive hazard mitigation and personalized safety recommendations (Kumar, P., & Kaur, M., 2021). The integration of AR could provide miners with real-time, context-aware information and guidance, enhancing situational awareness and decision-making capabilities (Zhang, L., & Yang, Z., 2016). Additionally, the integration of Smart Helmets with robotics and autonomous systems could enable more efficient remote monitoring and control of mining operations, reducing the need for human presence in hazardous environments (Li, Q., Liu, H., & Ning, H., 2017).

### 9.2 Improved Sensor Technology

Continued advancements in sensor technology will likely result in the development of more accurate, reliable, and energy-efficient sensors for IoT-based Smart Helmets. This could include the development of novel sensors for detecting previously unmonitored hazards, as well as improvements in the performance and longevity of existing sensors (Saha, H. N., & Mamun, K. A., 2016). Research in this area could also focus on the development of self-calibrating or self-healing sensors, reducing the need for manual maintenance and calibration (Kumar, P., & Kaur, M., 2021).

### 9.3 Enhanced Wireless Communication

As wireless communication technologies continue to evolve, future IoT-based Smart Helmets may benefit from improved connectivity, reliability, and data transmission rates. This could include the adoption of emerging communication protocols, such as 5G or Low Earth Orbit (LEO) satellite networks, which offer increased bandwidth, lower latency, and better coverage in remote or challenging environments (Li, Q., Liu, H., & Ning, H., 2017). Research in this area could also focus on the development of more resilient and adaptive communication systems, capable of maintaining connectivity in the face of interference or changing environmental conditions (Zhang, L., & Yang, Z., 2016).

### 9.4 Energy Harvesting and Improved Battery Life

Future IoT-based Smart Helmets may incorporate energy harvesting technologies, such as solar, thermal, or piezoelectric energy harvesting, to supplement or replace traditional battery systems (Saha, H. N., & Mamun, K. A., 2016). This could extend the operational lifetime of the Smart Helmets and reduce the need for battery replacement or recharging, contributing to increased sustainability and reduced maintenance costs (Kumar, P., & Kaur, M., 2021).

## 10. Conclusion

IoT-based Smart Helmets have the potential to revolutionize hazard detection and monitoring in the mining industry, improving worker safety, increasing operational efficiency, and reducing costs. By incorporating various sensors and communication technologies, Smart Helmets can detect and alert miners to the presence of hazardous conditions, such as toxic gases, high temperatures, excessive noise levels, and impacts, in real-time. The data collected by the Smart Helmets can also be used for real-time monitoring, decision-making, and regulatory compliance.

However, the deployment of IoT-based Smart Helmets also presents challenges and limitations, including technology adoption, sensor performance, wireless communication, and data security and privacy, and battery life. Addressing these challenges requires a thorough understanding of the specific mining environment, careful selection of sensors and communication protocols, robust security measures, and ongoing maintenance and optimization of the system.

In summary, IoT-based Smart Helmets have the potential to significantly enhance hazard detection and monitoring in the mining industry, contributing to safer and more efficient mining operations. As technology continues to advance and the challenges and limitations are addressed, IoT-based Smart Helmets are likely to become an increasingly integral part of the mining industry's safety practices and procedures.